# COST CONSIDERATION AND A POSSIBLE CONSTRUCTION TIMELINE OF THE CEPC-SPPC

W. Chou[#], Fermilab, Batavia, IL 60510, U.S.A.

*Abstract*

This paper discusses the cost consideration and a possible construction timeline of the CEPC-SPPC study based on a preliminary conceptual design that is being carried out at the Institute of High Energy Physics (IHEP) in China.

## INTRODUCTION

The discovery of the Higgs boson in 2012 at CERN was not only a milestone in particle physics, but also a trigger in the world high-energy physics (HEP) strategic planning for a renewed interest in future circular colliders. Because the Higgs mass is low (125 GeV), a circular *e+e-* collider can be built to serve as a Higgs factory. But the ring size must be big in order to combat the synchrotron radiation problem. Such a large size ring would be ideal to house a *pp* collider with an energy much higher than that of the LHC. Based on this consideration, the IHEP proposed to build a 50-100 km ring in China. It would first be used as a Higgs factory with the name Circular Electron-Positron Collider (CEPC), then as a 70-100 TeV Super Proton-Proton Collider (SPPC).

A preliminary conceptual design study of the CEPC-SPPC started in earnest in early 2014. In order to be considered as a line item listed in the Chinese government's next Five-Year Plan (2016-2020), the study was put on a fast track – a preliminary conceptual design report is due the end of 2014.

A general description of the CEPC-SPPC can be found in another presentation at this workshop [1]. This paper will discuss the cost consideration and a possible construction timeline of a 50 km ring.

Table 1 and 2 list the top level parameters of the CEPC and SPPC, respectively. Please note that the luminosity of the SPPC has not yet specified because there is an ongoing debate in the world HEP community about the required luminosity of a future high energy *pp* collider [2].

Table 1: Top Level Parameters for CEPC.

| Parameter | Design Goal |
|---|---|
| Particles | *e+, e-* |
| Center of mass energy | 240 GeV |
| Integrated luminosity (per IP per year) | 250 fb$^{-1}$ |
| No. of IPs | 2 |

Table 2: Top Level Parameters for SPPC.

| Parameter | Design Goal |
|---|---|
| Particles | *p, p* |
| Center of mass energy | 70 TeV |
| Integrated luminosity (per IP per year) | (TBD) |
| No. of IPs | 2 |

Figure 1 is a layout of the CEPC. The circumference is about 54 km. There are 8 arcs and 8 straight sections. Four straight sections, about 1 km each, are for the interaction regions and RF; another four, about 800 m each, are for the RF, injection, beam dump, etc. The lengths of these straight sections are determined when the future need of large detectors and complex collimation systems of the SPPC are taken into account. The total length of the 8 straight sections is about 14% of the ring circumference, similar to the LHC. Among the four IPs, IP1 and IP2 will be used for *e+e-* collision, whereas IP2 and IP4 for *pp* collision.

The tunnel will be underground, about 50-100 m deep. It will accommodate three ring accelerators: the CEPC collider, the SPPC collider, and a full energy booster for the CEPC. Therefore, the tunnel must be big, about 6 m in width as shown in Figure 2. While the two colliders will sit on the floor, the booster will hang on the ceiling, similar to the Recycler in the Main Injector tunnel at Fermilab.

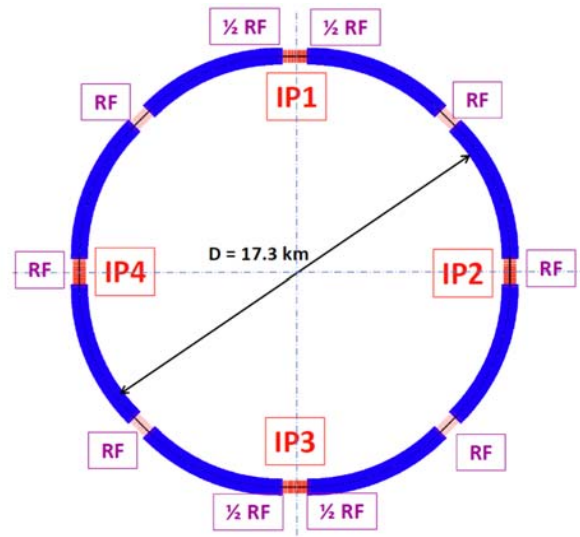

Figure 1: CEPC layout.

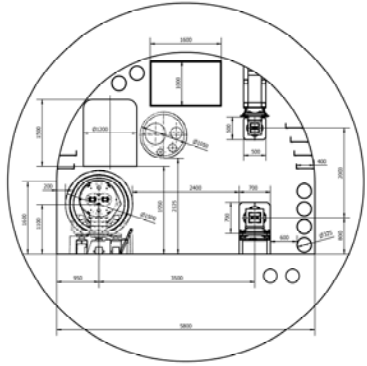

Figure 2: Tunnel cross section. The magnet on the left is the superconducting magnet of the SPPC, the magnets on the right are that for the CEPC collider (bottom) and the Booster (top), respectively. The tunnel width is about 6 m.

## COST CONSIDERATION

The synchrotron radiation power of the $e+$ and $e-$ beam is 50 MW each, which must be provided by the RF, Therefore, the most expensive technical systems of the CEPC are: (1) the superconducting RF (SRF) system; (2) the RF power source; and (3) the cryogenic system.

There are two SRF systems:

- 1.3 GHz 9-cell cavities for the booster, similar to the ILC, XFEL and LCLS-II
- 650 MHz 5-cell cavities for the CEPC collider, similar to the ADS and PIP-II

The synergy makes it possible to have a reliable cost estimate based on experiences of the other accelerators,

Two cost references were particularly useful: the actual cost of LEP1 and LEP2, and the cost estimate of the LCLS-II 4 GeV SRF linac.

The LEP1 cost was well documented [3,4]. The total in 1986 prices was 1.3 billion Swiss francs (BCHF). LEP2 added 288 SRF in the 1990s for about 0.5 BCHF [5]. Taking into account the inflation, the construction of LEP1 and LEP2 would cost roughly 2.6 BCHF in today's prices. As the CEPC is twice as large as the LEP, plus a full-energy booster, the cost would be about 7 BCHF if it is built in Switzerland. But the cost in China is lower, especially the civil construction. The goal is to reduce it by half to about 3.5 BCHF, or 20 billion Chinese Yuan.

But, of course, a simple cost scaling will not work. For example, while the civil construction in China can be much cheaper than in Switzerland, the klystron price is the same around the world as only a few vendors can make them.

Two cost estimate exercises were carried out at the IHEP: one by the magnet group, another by the vacuum group. Each group was given the LEP design and was asked to estimate the cost if the identical magnet or vacuum system was built in China. The result showed that the LEP magnet would cost 30% less if fabricated in China. But the saving on the vacuum was smaller because China does not have the advanced aluminium extrusion technology.

The LCLS-II is another useful reference. Its 4 GeV linac uses the 1.3 GHz 9-cell ILC type cavities and cryomodules. The cost is 2.7 million US dollars (USD) per module, or a total of 105M USD for 38 modules. But this cost does not include non-superconducting RF part (klystron, modulator, RF distribution, etc.). The CEPC booster needs 32 cryomodules (1.3 GHz), and the collider 96 cryomodules (650 MHz).

Several measures for cost reduction were taken:

- The guideline is: if there are several options for technology, the cheaper one is chosen for the baseline.
- Two beams ($e+$ and $e-$) will share the same beam pipe as in the LEP and CESR.
- As the CEPC has smaller beam emittance than the LEP, the magnet aperture is reduced by 20%, which saves the construction as well as the operation cost.
- Although solid state is more reliable and easier to maintain than the klystron, the latter is cheaper and has higher efficiency. So klystron is chosen for the baseline.
- For tunnel construction, the tunnel boring machine (TBM) is faster than the explosion method. But the latter can save 20-30% cost and is thus chosen. Moreover, this method can make a city gate-shape tunnel as in Figure 2, which gives more usable space than a circular shape.

A Work Breakdown Structure (WBS) was used for cost estimate. Figure 3 shows the relative cost of each system (excluding the civil, which is under study).

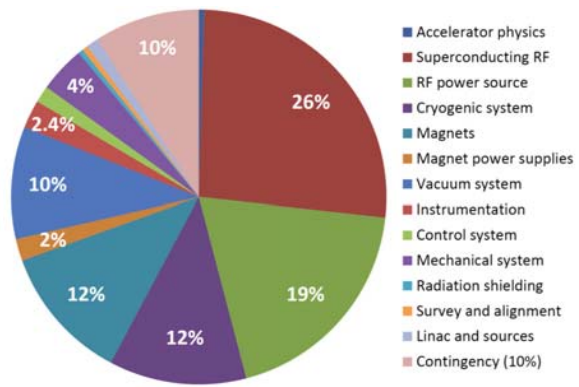

Figure 3: Relative cost of the CEPC technical systems.

It should be pointed out that the cost of the cryogenic system in this plot (12%) is based on a high efficiency HOM damper, which needs to be developed. Both the 1.3 GHz and 650 MHz SRF will operate at 2° K. The Carnot efficiency from the ILS study is listed in Table 2. Because the average beam current in the CEPC is high (16.6 mA for each beam), HOM loss in the cavity is significant (2.3 kW per beam in each cavity). Most of the HOM power must be taken out and dissipate at higher temperatures. Table 4 is the required efficiency of the HOM damper, which is very demanding. How to design and implement such a damper is a critical R&D for the CEPC.

Table 3: Carnot Efficiency for CEPC SRF.

|  | 40 K to 80 K | 5 K to 8 K | 2K |
|---|---|---|---|
| Efficiency in W/W | 16.4 | 197.9 | 703.0 |

Table 4: Required Efficiency of the HOM Damper.

|  | 40 k to 80 k | 5 k to 8 k | 2k |
|---|---|---|---|
| HOM heat load distribution | 3% | 0.3% | 0.1% |

Power consumption determines the operation cost. When the Tevatron was running, the average total power usage at Fermilab was 58 MW. When the LHC was running, CERN used 183 MW (average over 2012). A consensus for a future circular Higgs factory is that the power should not exceed 300 MW, in which 100 MW is for synchrotron radiation. In other words, the wall plug efficiency (the ratio between the beam power and the wall power) should be 33%. This is a tall order as today's most efficient accelerator, the PSI cyclotron, has only an efficiency of 18%. The design efficiency for the ILC is just 9.6%. In order to have a highly efficient CEPC, one needs a highly efficient SRF system. The recent development of a new type of klystron (Collector Potential Depression, or CPD) is of particular interest as its claimed efficiency can be as high as 80%. Reuse and recycle of waste power from the accelerator is part of a general study nicknamed "green accelerators." Figure 4 shows the relative power consumption of each system in the CEPC.

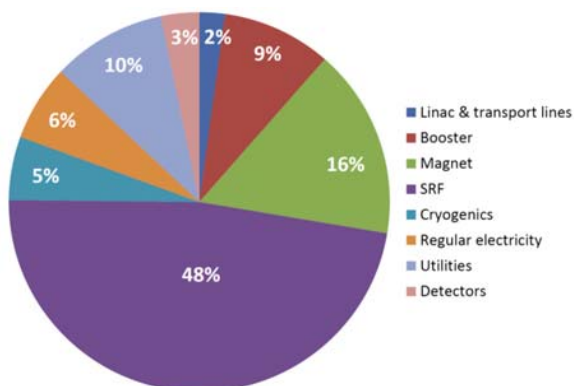

Figure 4: Relative power consumption of each system in the CEPC.

## A POSSIBLE TIMELINE

Figure 5 shows a presently conceived timeline. It consists of following stages:

- The first milestone is to complete a Preliminary Conceptual Design Report (Pre-CDR) by the end of this year. It will be used to apply for R&D fund for the next five years.
- The Chinese government's 13[th] Five-Year Plan starts in 2016. If the CEPC gets approval, the R&D will take place 2016-2020.
- The construction will (hopefully) start in 2021 in the government's 14[th] Five-Year Plan and take about 8 years.
- The experiment can start in 2028 during the 15[th] Five-Year Plan.
- For the SPPC, the focus is to develop cost effective high field superconducting magnet (16-20 Tesla) by means of $Nb_3Sn$ and HTS superconductors. This will take 15 years. The engineering design will start in 2030 and the construction to start around 2035.

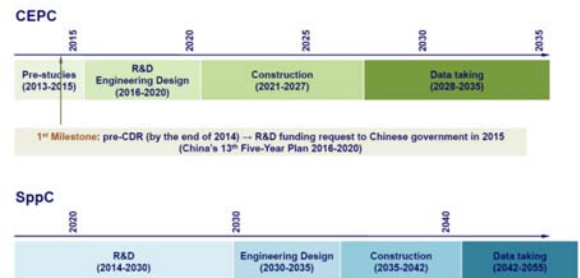

Figure 5: A possible timeline.

Of course the realization of such a fast track timeline depends on many factors. Some are under our control, some are not. At this moment, the effort is focused on meeting the first milestone, namely, to complete the Pre-CDR in the next several months.

A critical path of the CEPC timeline is to have a successful R&D for the two SRF systems:

- Collider: 650 MHz, 384 cavities in 96 cryomodules;
- Booster: 1.3 GHz, 256 cavities in 32 cryomodules.

This would be the largest SRF installation in the world. To succeed with designing, fabricating, commissioning and installation of such a system, a significant investment in R&D, infrastructure and personnel is necessary. The R&D has two parts:

- Prototyping as well as technology development for several critical components, in particular, the power coupler and the HOM damper.
- Pre-series production:
  - 15-20 1.3 GHz cavities and 30-35 650 MHz cavities
  - A large RF facility similar to that at Jlab, Fermilab and DESY for cavity inspection and tuning set ups, RF lab, several vertical test stands, clean rooms, HPR systems, FPC preparation and conditioning facility, cryomodule assembly lines, horizontal test stations, high power RF equipment, a cryogenic plant, etc.
- Capable to assemble 1 Booster modules and 2 Collider module each month
- To have at least two vendors for each type of RF
- Personnel development

This R&D plan will absorb enormous resources and take a number of years. If the construction starts in 2021, the tunnel will take 4-5 years to finish. So there should be enough time for the pre-series production to complete before mass production.

# REFERENCES

[1] Q. Qin, "Overview of the CEPC Accelerator", Proceedings of this workshop.
[2] B. Richter, "High Energy Colliding Beams – What is the Future?", to appear in the *Reviews of Accelerator Science and Technology*, Volume 7, World Scientific (2014).
[3] "Financial Position of the LEP Project – Final Report," CERN/FC/3313, May 30, 1990.
[4] H. Schopper, "LEP – The Lord of the Collider Rings at CERN 1980-2000," Springer (2009).
[5] S, Myers, private communication.